\documentclass[conference]{IEEEtran}

\IEEEoverridecommandlockouts
\usepackage{cite}
\usepackage{amsmath,amssymb,amsfonts}
\usepackage{algorithmic}
\usepackage{graphicx}
\usepackage{textcomp}
\usepackage{xcolor}
\usepackage{placeins}
\usepackage{fancyhdr}
\def\BibTeX{{\rm B\kern-.05em{\sc i\kern-.025em b}\kern-.08em
    T\kern-.1667em\lower.7ex\hbox{E}\kern-.125emX}}

\fancypagestyle{firstpage}{%
 \fancyhf{}%
 \fancyhead[L]{2024 IEEE International Conference on Signal Processing, Information, Communication and Systems (SPICSCON)\newline
  {1-2 November, Khulna University, Khulna, Bangladesh}}
  
 \fancyfoot[L]{%
  \normalsize{979-8-3315-1021-3/24/\$31.00 ~\copyright2024 IEEE}
 }%
}

\begin{document}
\IEEEoverridecommandlockouts

\title{MODELING THE SHARING AND DIFFUSION OF FAKE
NEWS IN SOCIAL MEDIA\\
%{\footnotesize \textsuperscript{*}Note: Sub-titles are not captured in Xplore and
%should not be used}
%\thanks{Identify applicable funding agency here. If none, delete this.}
}

% \author{\IEEEauthorblockN{Umme Faria Moon}
% \IEEEauthorblockA{\textit{Computer Science and Engineering} \\
% \textit{Jahangirnagar University}\\
% Dhaka, Bangladesh \\
% moon.stu20181@juniv.edu}
% \and
% \IEEEauthorblockN{MD Ahsan Habib Rasel}
% \IEEEauthorblockA{\textit{Computer Science and Engineering} \\
% \textit{Jahangirnagar University}\\
% Dhaka, Bangladesh \\
% ahsanhabib.stu2018@juniv.edu}
% \and
% \IEEEauthorblockN{Md. Musfique Anwar}
% \IEEEauthorblockA{\textit{Computer Science and Engineering} \\
% \textit{Jahangirnagar University}\\
% Dhaka, Bangladesh \\
% manwar@juniv.edu}
% }
\author{
    Umme Faria Moon, MD Ahsan Habib Rasel, Md. Musfique Anwar \\
  Department of Computer Science and Engineering\\
    Jahangirnagar University \\
    Dhaka, Bangladesh \\
    moon.stu20181@juniv.edu, ahsanhabib.stu2018@juniv.edu, manwar@juniv.edu
}

\maketitle
\thispagestyle{firstpage}
\begin{abstract}
The use of social media platforms has been gradually increasing and fake news spreading is becoming an alarming issue nowadays. The spreading of fake news means disseminating false, confusing, and spurious information which hurts families, communities etc. As a result, this issue has to be resolved sooner so that we can limit the spread of fake news in the virtual world. One needs to identify the fake news spreader to address this issue. In this research, we have tried to reveal the users who are most likely to share fake news as well as the spread prediction that shared pieces of fake news in the social network. We take into account the users’ information, such as follower counts, like counts, and retweet counts along with
users’ topical interests on different topics as well as connection strength by considering the follower-following ratio. We also consider the complexity features, stylistic features, and psychological effects of news. Finally, we applied different machine-learning algorithms to evaluate the performance of the proposed model. Our observation is that the probability of spreading a piece of news shared by users having more followers as well as more likes and retweet counts (aka \textit{influential} users) is higher compared with other users.

\end{abstract}

\begin{IEEEkeywords}
Social Media, Fake News, Spreader, Topical Interest, Diffusion.
\end{IEEEkeywords}

\section{Introduction}
Nowadays, online social networks (OSNs) (e.g., Facebook, Twitter, Google+, etc.) are very popular platforms for communication, entertainment, and broadcasting news among users\cite{Anwar,Anwar_ADC}. Besides these advantages, the major drawbacks of social media are the spreading of misinformation and privacy issues. Social media also places itself as the main channel for the spread of fake news due to the simplicity of sharing facts and data. Through the similarity in the false stories, one story can be passed on rapidly from one user to many audiences within seconds. This rapid diffusion of fake news poses a major problem, as it can lead to the spread of misinformation, confusion, and even harm. In our research, we have tried to show how false news can spread quickly on social media. We have considered Twitter (current name: "X") as a social media platform and are thinking about some important features for this purpose.

Twitter is an essential platform for the investigation of fake news diffusion. Users can share breaking news, viral content, and other information almost instantaneously, often without verifying the accuracy of the claims\cite{Badhan, Sarmistha,Anwar_Wise}. This quick-sharing behavior can contribute to the rapid diffusion of fake news, as users may be more inclined to share sensational or attention-grabbing content without scrutinizing its validity. Thus, understanding of users' private behaviors and the factors that lead to fake news spreading on Twitter is a gap that still needs to be bridged. 

Fake news can be defined as false news, misinformation, or rumors. Generally, fake news has diffused rapidly via social media platforms in recent times. In our research, we focused on the diffusion of this fake news in Twitter. We investigated the role of individual users in spreading fake news on Twitter. We paid attention on the number of tweets each user shares, as well as the number of retweets their followers generate. Furthermore, we also considered the content of these tweets to identify those as fake or real news. This allows us to quantify the extent to which each user is contributing to the spread of fake news. 

%We also used an AUROC(Area Under the Receiver Operating Characteristics) and 5-fold cross-validation for assessment.

%Thus in this paper, we have tried to express the Fake News diffusion during the pandemic time. 

The main objective of our research is to build a machine-learning approach that is capable of observing and analyzing the dissemination of fake information by identifying the main sources, potential factors, and interesting patterns correlated to the spread of fake news spreading in OSNs. We applied topic modeling technique such as T-LDA (Twitter LDA) to find users' topical interests and then applied machine learning algorithms such as Random Forest, XGBoost, HistGradientBoostingClassifier, Logistic Regression, SVM, and Decision Tree to classify social content as fake or real. We also used Random Forest Regressor and XGBoost Regressor to measure the spread or diffusion of fake news.

%We want to stop the spread of fake news. Our model will track the origin of a piece of fake news and show how it spreads. Besides, our purpose is to mark the misleading information and raise public awareness about it.

\section{Related Work}

We have studied some relevant research works. Heru et al.~\cite{b1} classify fake Twitter accounts and genuine Twitter accounts to discover fake news spreaders. Joy et al.~\cite{b2} considered ICM (Independent Cascade Model) and LTM (Linear Threshold Model) as the baselines to develop a model that characterized fake news sharing in social media. Jisu Kim et al.~\cite{b3} assessed a dataset that shows an extensive diffusion of fake news during the time of COVID-19. They produce a dataset named FibVID (Fake news information-broadcasting of COVID-19). Yaqub et al.~\cite{b4} showed the effect of credibility indicators on people's intention to share news where, credibility indicators are the pointers that assist a user in the truthfulness or reliability of something. \\

Recently, researchers applied machine learning techniques to identify fake social media content. For example, M. Sudhakar et al. \cite{b5} applied the SVM model to detect fake news regarding COVID-19. %from social media using SVM. There, they used the  dataset which contains about 1,375,592. Then they applied both Machine learning and Deep Learning models. 
Gulzar et al.\cite{b6} detect Bangla fake news using SVM and MNB (Multinomial Naive Bayes) classifier. %MNB recognized Bangla fake news. Then they applied SVM and found an accuracy of 96.64 percent.  
R. Jehad et al.\cite{b7} classify fake news using the Random Forest model. Sudhakar et al.\cite{b8} evaluated the precision in the detection of fake news using the novel BERT (Bidirectional Encoder Representations for Transformers) technique. %and compared the performance with the Random Forest. \\
Finding out the fake news spreaders can also be an effective method of modeling the fake news diffusion. Anu et al.~\cite{b9} found that fake news spreaders consider linguistic features such as styles, words as well as personality features such as sentiment analysis.% These features are measured from a set of collected tweets. They also model the problem as a binary classification task. \\

%Previous works showed that the techniques of fake news detection mainly concentrate on the news content, user behavior, or fact-checking. However, utilizing social engagement is also very important to detect fake news. 
Shu et al.~\cite{b10} utilize the tri-relationship between the publisher, news, and appropriate user attachments in social media.
%Some computational Social Science observers explore agent-oriented models.
Nicholas et al.~\cite{b11} proposed a Cognitive Cascade model that refers to the circumstance where a single portion of information, behavior, or hypothesis spreads rapidly in OSNs. %  In this research, They propose a Cognitive Cascade model that acts like a network-based Diffusion Model as a Public Opinion Diffusion(POD) model. 
BILAL et al.\cite{b12} explored an emotional analysis of the misinformation and determined four main categories of false information such as propaganda, hoaxes, satires, and clickbait.% They analyzed a collection of real and fake tweets from different perspectives. %Another research by \cite{b13} showed an experimental comparison of fake news detection with different machine-learning models.\\

%Our research mainly focuses on how fake news spreads during a pandemic. We have applied T-LDA(Twitter LDA) for topic modeling and Some Machine Learning Approaches such as Random Forest, XGBoost, HistGradientBoostingClassifier, Logistic Regression, SVM, and Decision Tree for Classification. We also use 5-fold cross-validation to evaluate our model's performance.

\section{Dataset}\label{ch:dataset}
\subsection{Dataset Description}

We considered the dataset FibVID\footnote[1]{https://github.com/merry555/FibVID} (Fake News Information-Broadcasting Dataset of COVID-19) \cite{b3} which contains detailed information about real and fake news sharing associated with COVID and non-COVID news. %They utilized their dataset from the two widely used fact-checking sources, Politifact and Snopes to ensure its reliability. This dataset is generated to manipulate the misinformation spreading during the COVID-19 period. FibVID produces valuable perception propagation sequences of fake news, user characteristics, and user involvement in the diffusion of false news related to both COVID and non-COVID-based topics. 
This dataset consists of three segments: (i) news claim, (ii) claim propagation, and (iii) user information.%\cite{b3} This dataset revealed an analogy of the topic labeled by true and fake claims of the COVID and non-COVID issues.
\subsubsection{\textbf{News Claim}}
The news claim segment contains comprehensive information about different news items correlated with COVID and non-COVID topics. 
We created a feature extraction framework to measure the complexity analysis, psychological impact, and stylistic features of the news text.
\begin{itemize}
      \item \textit {Complexity feature:} We considered three key indicators for complexity analysis: smog index, lexical diversity, and average word length, thus dividing the news complexity into three categories as simple, medium, and complex.
      \item \textit{Psychological feature:} We considered two psychological aspects: sentiment polarity and sentiment subjectivity. Thus, we group the news content as positive, negative, or neutral based on the sentiment polarity.
      \item \textit{Stylistic Feature:} Considering writing style along with presence of personal and impersonal pronouns, the news content is categorized as non-stylic and stylic.
      
\end{itemize}
                    
\subsubsection{\textbf{Claim Propagation}} 
The claim propagation part of the dataset shows the propagation patterns of the news claims that diffuse on Twitter by assuming insights into the mechanics of misinformation spreading in OSNs. It includes data corresponding with the tweets and retweets connected to each news claim, tweet IDs, user information, post text, retweet counts, like counts, hashtags, and creation dates and they are categorized into COVID and non-COVID topics.

\subsubsection{\textbf{User Information}}
This section gives information about the users' characteristics involved in sharing fake news on social media. It includes user profile information like user IDs, descriptions, follower counts, following counts, and dates. %By analyzing user's behavior, researchers can recognize the users who are influenced to share misinformation on Twitter. 

\subsection{Dataset Analysis}
In this FibVID dataset, about 772 claims are related to COVID-19 topics. Among them, 230 claims are granted as true and 569 as fake. Besides, 581 claims are related to non-COVID, of which 150 claims are granted as true and 431 as fake. The dataset consists of 221,253 tweets posted by 144,741 users. The average number of each claim is different. In the dataset, COVID true claims have an average of 634.79 tweets and COVID fake claims have an average of 945.91 tweets. Additionally, non-COVID true claims have an average of 426.43 tweets and fake claims have an average of 590.00 tweets. %Besides, the dataset has claim propagation depth information, user involvement, and the equivalence between the original tweets and claims. %These values give a comprehensive description of the configuration and characteristics of the dataset.  

\section{Proposed System}
The common activities in OSNs are that users connect and % generate different content on different topics, share or like content if they feel interested, etc. 
 Retweeting or sharing is the fundamental appliance of information dissemination in OSNs as the original user-generated content is shared with a new set of audiences \cite{b14}. 

\iffalse
\subsection{Problem Description and Proposed Framework}
\label{sec:ch3probleTwitterm_description}
\subsubsection{Problem Description}

We first introduce several related concepts before formally formulate the problem of detection and diffusion of fake news in OSNs.

\vspace{0.5ex}

\noindent\textbf{Social Network:} A social network can be represented as a graph \textit{G = (U, E)}, where \textit{U} representing users and \textit{E} points the connections among the users. 

\vspace{0.5ex}

\noindent\textbf{Topic:} A topic is a distribution of a group of words. For example, \textit{algorithm} topic contains different words like sort, search, code, tree, etc.

\vspace{0.5ex}

\noindent\textbf{Topic distribution:} Each user can be interested in multiple topics. Formally, each user \textit{u} $\in$ \textit{U} is connected with a vector $\theta_{u} \in \mathbb{R}^T$ of \textit{T}-dimensional topic distribution ($\sum_{z} \theta_{uz}$ = 1). Each element $\theta_{uz}$ denotes the probability (interest) of the node (user) on topic \textit{z}.

\vspace{0.5ex}
\fi

%\FloatBarrier

%\textbf{Problem Definition:} If we think about a social graph \textit{G = (U, E, M)}, where \textit{U} indicates the set of users, \textit{E} indicates the set of edges, and \textit{M} is the set of all tweets/news published by all the users or the news sources. The task is then to detect whether recipient users will share the content or not thus measuring the spread of the content.

\subsection{Overview of the Proposed Framework}
In the proposed model, we consider how information flows from the sender of the new tweet/news through the network and how recipients process the information from incoming tweets and update their properties. Our proposed model has three steps as shown in Fig. \ref{fig:system_architecture}:

\begin{figure}
\centering
\includegraphics[width=0.9\columnwidth]%{system_architecture.jpg}
{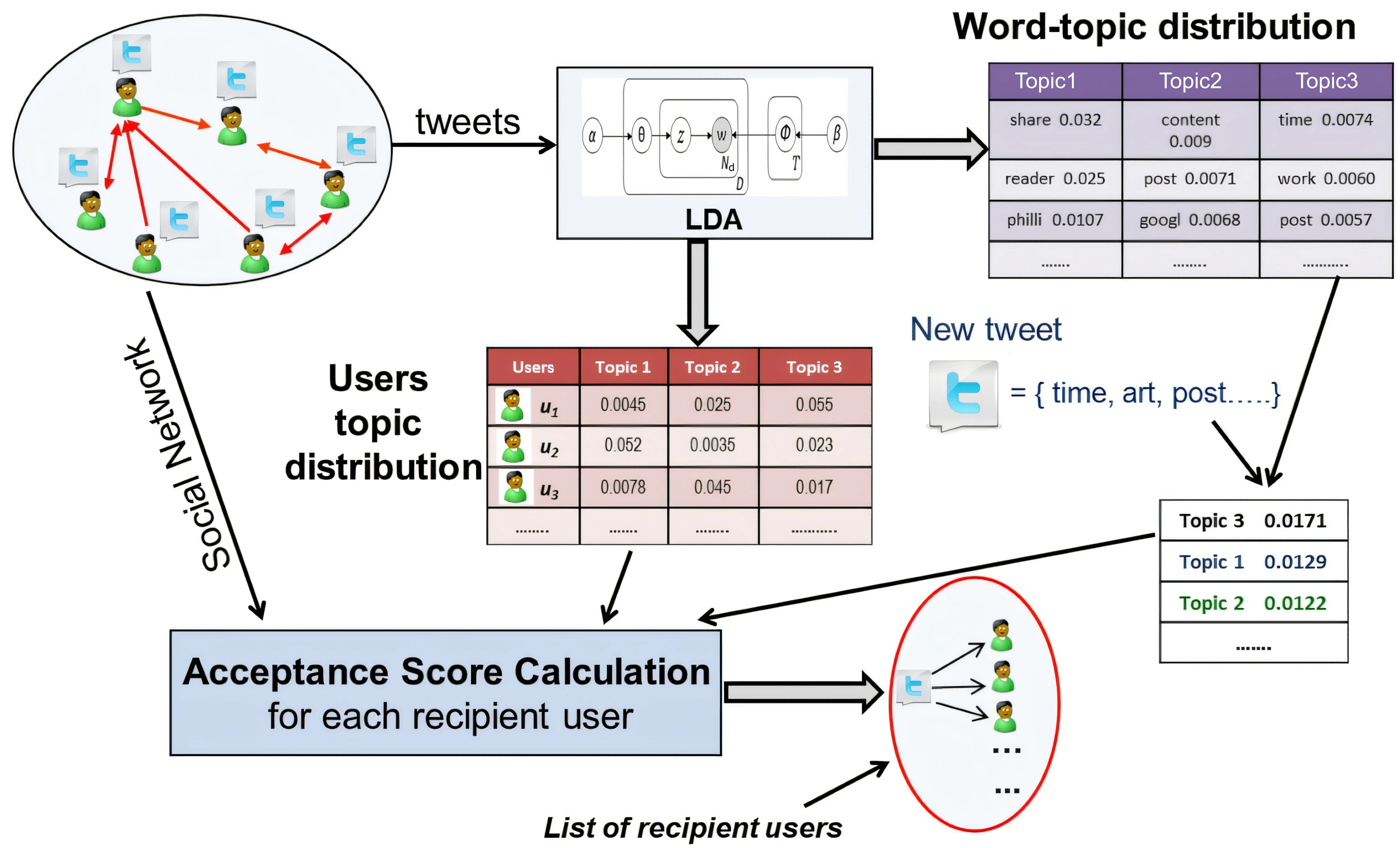}
\caption{Example of Tweet propagation} \label{fig:system_architecture}
\end{figure}

\textbf{Step 1:} At first, we need to find the latent topic distribution for all users. In our work, we applied Twitter-LDA (T-LDA) \cite{b16} as our topic modeling approach to initialize the topic distribution of each user. The input of this step is the tweets of all the users and the outcome is the user's topic distribution and word-topic distribution of each topic which contains the importance weight of each word in each topic.

\textbf{Step 2:} The input to the system will be composed of (tweets/news, users, and social connections). We measure the social strength by calculating \textit{follower-following} ratio (\textit{TFF}). 

\textbf{Step 3:} In the final step, we classify tweet as fake or real as well as estimate the future spread of the new tweet. To do so, we measure the acceptance score of each recipient user for the new tweet by considering the topic of the new tweet and the topical interests of the recipient user on those topic(s). %Multiple paths from the sender to the recipient may exist for the same tweet, each with its corresponding perceived information value. Recipient users will determine their evaluation of the information by incrementally merging all the pieces of value they receive for the same tweet from various paths. If the (accumulated) acceptance score is higher than a certain threshold value then we add that recipient user to the list of accepted users of $m$ who will forward his acceptance score to his neighbors. The input of this step is the topic distribution of the users that we found from Step 1 and the TFF scores from Step 2. The final outcome is the expected list of accepted users who have the new tweet.

\iffalse
\subsection{Users Topic Distribution}\label{sec:ch3LDA}

Topic modeling is the task of identifying which underlying concepts are discussed within a collection of documents, and determining which topics each document is addressing. We applied Twitter-LDA\cite{b16} on the tweets to learn the latent topic distribution of users. The idea is that the documents consist of different random topics. T-LDA is a generative process for generating each document as follows (Figure ~\ref{fig:LDA_Model}):
\begin{figure}[htbp]
\centering
\includegraphics[width=0.9\columnwidth]{LDA_Model.jpg}
\caption{Graphical Representation of LDA Model} \label{fig:LDA_Model}
\end{figure}

\begin{enumerate}
    \item Consider a random distribution of topics.
		\item For every word in each document: \begin{enumerate}
        \item Randomly choose a topic from the distribution over topics in step (i).
        \item Randomly choose a word from the corresponding distribution over the words associated with the chosen topic.
    \end{enumerate}
    \item The process is iterated for all the words in the document.
\end{enumerate}

\fi

\subsection{Process Overview}
%In our research, first, we collected our dataset from FibVID and we processed it by feature extraction. Then we use T-LDA for topic modeling and we find out the topics based on used user interest. Then we calculate TFF to find out user's engagement level and social influence. Then we applied some appropriate machine-learning algorithms and trained our model. After that, we obtain the result by measuring the Accuracy, Precision, Recall, AUROC, F1-Score , RMSE and R$^2$ value

\subsubsection{Processing Dataset}

%We conducted a broad analysis to learn what the users like and what features of the news articles are more interesting to them. 
We first extracted unique tweet texts from the dataset and applied T-LDA to identify the topics of interest for each user. %The generative statistical model (LDA) allowed us to identify the underlying topics within tweets, providing insights into user preferences and engagement patterns.
%Later on, we proceeded to the text-reading stage of the news articles. 
Next, We created a feature extraction framework to measure the complexity analysis, psychological impact, and stylistic feature of the news text. We introduced Twitter Follower Following Ratio (TFF) to better understand the dynamics of content sharing on Twitter: 

\begin{equation}TFF = \frac{(\# \text{Follower} + 1)}{(\# \text{Followee} + 1)} \end{equation}

The TFF is calculated as the ratio of a user's followers to the following count, and we adjusted it to avoid division by zero by adding 1 to both the follower and the following count. This metric provides a measure of a user's influence and engagement level on the platform, which could impact the likelihood of sharing news content.

\begin{figure}
\includegraphics[width=0.9\columnwidth]{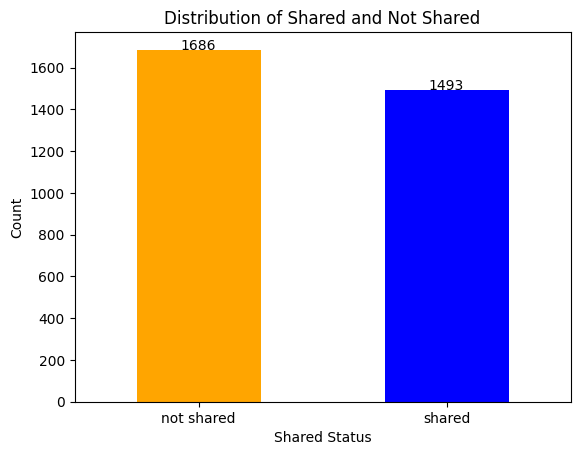}
\caption{Comparison of Classifier Performance Metrics} 
\vspace{-2ex}
\label{fig:shared_notshared}
\end{figure}

We carried out a labeling procedure to classify tweets as either ``shared" or ``not shared". The classification was made by finding whether the follower of the parent user has shared it or not. %We also tried to separate the tweets that got a lot of attention from those that did not, thus the binary result of each tweet in our data was obtained. 
Then we merged all the relevant features. After that, we eliminated the ones with incomplete or missing information and obtained 3,179 rows in the final dataset. Fig.~\ref{fig:shared_notshared} shows the distribution of ``shared" and ``not shared" labels within this dataset. The distribution shows us that the dataset is fairly distributed, with a slight majority of the tweets not being shared according to our set of thresholds. %This balance is essential for the following examination since it enables a more detailed exploration of the factors that are the reasons behind content sharing on Twitter.

\begin{figure}
\centering
\includegraphics[width=0.8\columnwidth]{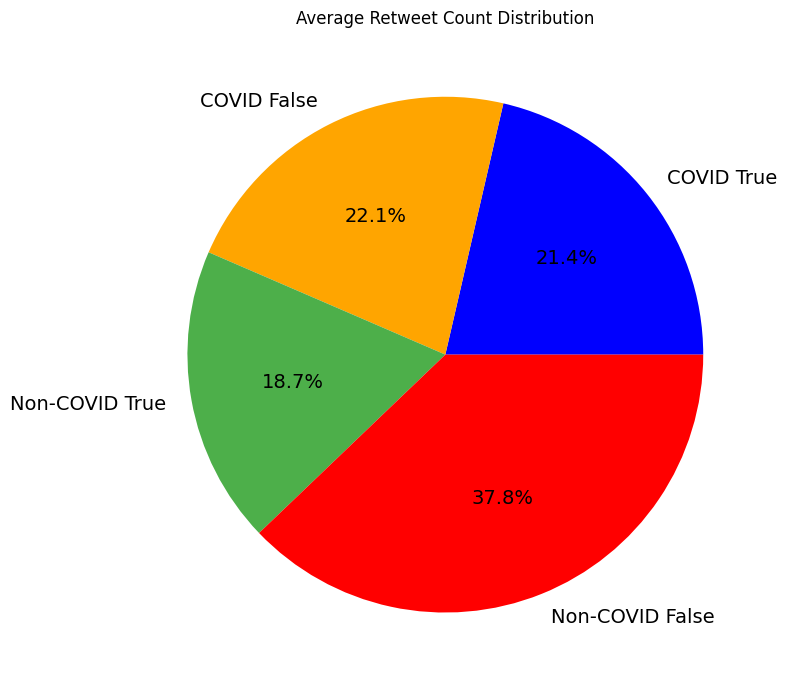}
\caption{Average retweet in each category} \label{fig:avg_retweet}
\end{figure}
%%\FloatBarrier

\subsection{Model Training}

\textbf{Classification: Predicting Fake News Sharing}
%\vspace{10pt}

We used supervised machine-learning algorithms such as Random Forest, XGBoost, HistGradientBoostingClassifier, Logistic Regression, SVM, and Decision Tree and trained our model with important features. We calculate the accuracy, precision, recall, and F1 score for our classification analysis.%We have removed unnecessary features by feature extraction.

%\vspace{10pt}
\textbf{Regression: Prediction of Retweet Count}

%\vspace{10pt}
We used both Random Forest Regressor and XGBoost Regressor to predict the number of retweets of news when a user posts it. %XGBoost Regressor is one of the most effective machine-learning approaches for developing a regression model. 
RMSE (Root of Mean Squared Error) and R$^2$ Score matrices are used for evaluation. 

%\subsection{Model Evaluation}
%We have evaluated our results with the confusion metrics of our models. We calculate the accuracy, precision, recall, and F1 score for our classification analysis. For regression analysis, we have used the R$^2$ Score and RMSE. We also consider AUROC (Area Under the Roc Curve) for evaluation. We have calculated an AUROC considering ICM and LTM as baselines.

\section{Experiment and Results}
%To predict whether a user will share the news or not, we have trained different classifier models with our processed dataset. We also predicted the number of users who will retweet a news post when that particular user shares it. We consider user information like the user's follower following count, and connection with other users for predicting this.
This section presents the performance of different machine learning algorithms regarding sharing fake news. 

\subsubsection{Classification} Table.~\ref{tab:my_label} shows the performance metrics of different the models. We have gained the highest accuracy from the Random Forest classifier, which is 92.05\%. The highest precision we have gained is 87.06\% from the XgBoost classifier whereas recall is 97.38\% obtained using the Random Forest classifier. Random Forest classifier also obtained best F1-Score of 92.05\%.

\begin{figure}%[htbp]
\includegraphics[width=0.9\columnwidth]{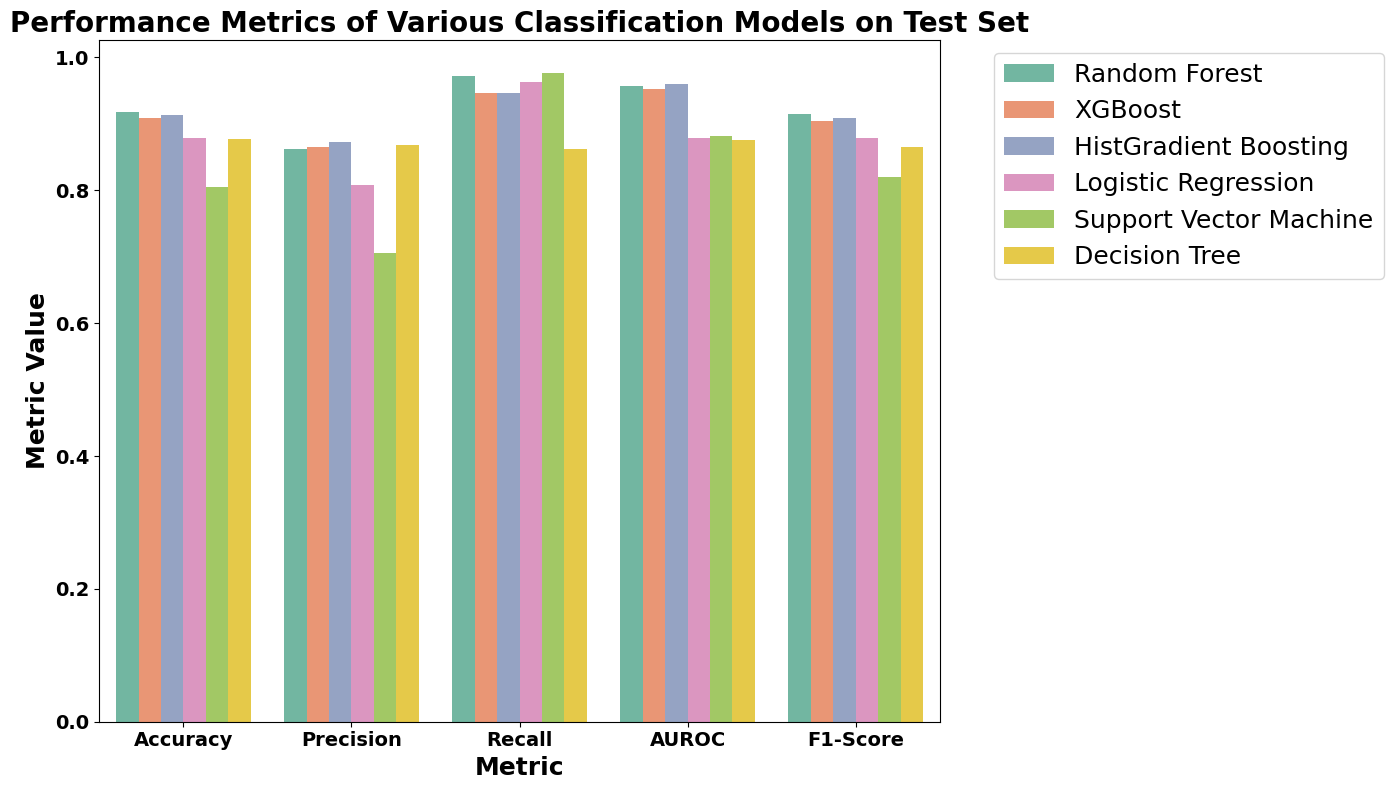}
\caption{Comparison of Classifier Performance Metrics} 
\vspace{-3ex}
\label{fig:comparison}
\end{figure}
%\FloatBarrier

\begin{figure}%[htbp]
\includegraphics[width=0.9\columnwidth]{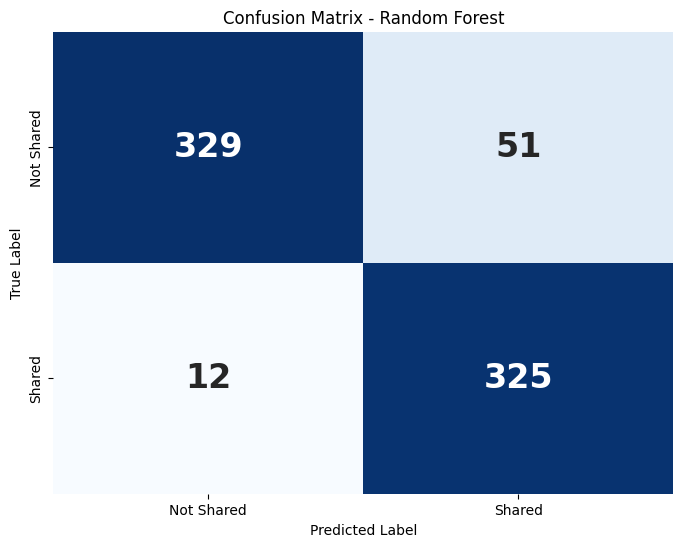}
\vspace{-1ex}
\caption{Random Forest Confusion Matrix} 
\vspace{-2ex}
\label{fig:random_cm}
\end{figure}

\begin{table}%[h]
    \centering
    \caption{Performance Metrics of Various Classification Models}
    \resizebox{\columnwidth}{!}{
        \begin{tabular}{|c|c|c|c|c|c|}
            \hline
            Model & Accuracy & Precision & Recall & AUROC & F1-Score\\
            \hline
            \hline
            Random Forest & \textbf{92.05} & 86.84 & \textbf{97.92} & \textbf{92.38} & \textbf{92.05} \\
            \hline
            XGBoost & 91.07 & \textbf{87.60} & 94.36 & 91.26 & 90.86 \\
            \hline
            HistGradient Boosting & 91.07 & \textbf{87.60} & 94.36 & 91.26 & 90.86 \\
            \hline
            Logistic Regression & 87.73 & 81.36 & 95.85 & 88.19 & 88.01 \\
            \hline
            SVM & 84.24 & 75.81 & 97.63 & 84.99 & 85.34 \\
            \hline
            Decision Tree  & 87.87 & 87.43 & 86.65 & 87.80 & 87.03 \\
            \hline
        \end{tabular}
    }
    \vspace{-5ex}
    \label{tab:my_label}
\end{table}

\begin{figure}%[htbp]
\vspace{-4ex}
\includegraphics[width=0.9\columnwidth]{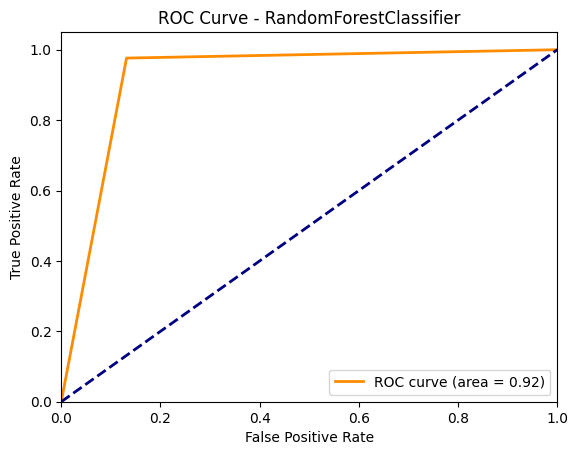}
\caption{Random Forest Roc Curve} \label{fig:random_roc}
\vspace{-2ex}
\end{figure}

\subsubsection{Regression} We have used two regression models to predict the number of retweets of news. From the result (see Table.~\ref{tab:reg_analysis}) we find that the XGBRegressor model performs better than the Random Forest Regressor. XGBRegressor has obtained an R$^2$ Score of 0.86 and an RMSE value of 3309.18.

\begin{table}%[ht]
\centering
\caption{Comparison of Regression Models}
\label{tab:regression_models}
\begin{tabular}{|l|c|c|}
\hline
\textbf{Model} & \textbf{RMSE} & \textbf{R$^2$ Value} \\
\hline
XGBRegressor & \textbf{3309.18} & \textbf{0.86}\\
\hline
Random Forest Regressor & 3618.11 & 0.84 \\
\hline
\end{tabular}
   
\label{tab:reg_analysis}
\end{table}

%The XGBRegressor model has given the best R$^2$ Score which is 0.76 and also RMSE value of 3309.18.

%The Random Forest Regressor has also done well. It has achieved  R$^2$ Score of 0.84.
%%\FloatBarrier

\begin{figure}%[htbp]
\includegraphics[width=0.9\columnwidth]{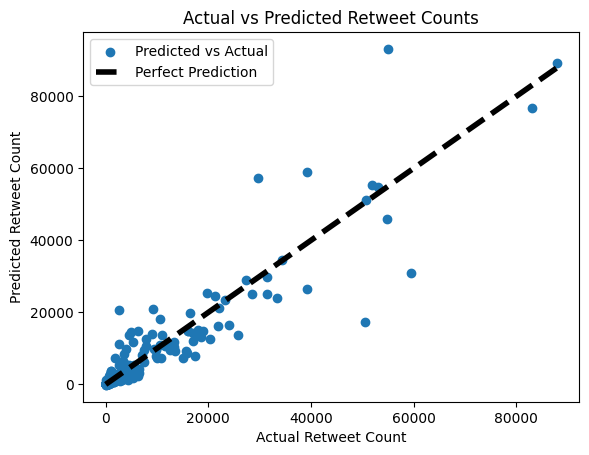}
\vspace{-2ex}
\caption{XGBRegressor:Actual retweet count vs Predicted retweet count } \label{fig:dt_pr}
\end{figure}

%%\FloatBarrier

\section{Conclusion}
Nowadays, stopping fake news diffusion is an important task. In this research, we identified the potential fake news spreaders in OSNs. We used T-LDA to extract topics from the news and applied six different classifiers. The performance of the Random Forest classifier exceeds all other classifiers with an F1-Score value of 92.05\%. %We also consider the AUROC for performance measurement. Random Tree classifier achieved 92.38\% of AUROC. All other performance measurements also show that the Random Forest classifier performs better among all the classifiers we used. We also predicted the number of users who will retweet a news post when that particular user shares it. We found that influential users have a higher retweet count than others. Influential users are those who have a higher number of followers. We consider the TFF(Twitter Follwer Following Ration) count for finding influential users. For this prediction, we used two regression models. The $R^2$ value for the Random Forest Regressor is 0.84 and for the XGB Regressor, it is 0.86. XGBoost Regression has performed better here. XGBoost Regression performs better than Random Forest regression. The RMSE count is also better for XGBoost regression. In this research, we have shown the spread of fake news. 
Future work is to expand it by analyzing multiple levels of a network and show how far a particular news can spread. %We aim to build an advanced real-time-based diffusion analysis model to consider the dynamic spread of news as it evolves. 

\vspace{12pt}

\end{document}